**Title: Infectious disease surveillance needs for the United States: lessons from COVID-19**

**Authors:** Marc Lipsitch[1,2], Mary T. Bassett[3], John S. Brownstein[4], Paul Elliott[5], David Eyre[6], M. Kate Grabowski[7], James A. Hay[6], Michael Johansson[8], Stephen M. Kissler[9], Daniel B. Larremore[9,10], Jennifer Layden[11], Justin Lessler[12], Ruth Lynfield[13], Duncan MacCannell[14], Lawrence C. Madoff[15], C. Jessica E. Metcalf[16], Lauren A. Meyers[17], Sylvia K. Ofori[2], Celia Quinn[18], Ana I. Ramos Bento[19], Nick Reich[20], Steven Riley[21], Roni Rosenfeld[22], Matthew H. Samore[23], Rangarajan Sampath[24], Rachel B. Slayton[25], David L. Swerdlow[2], Shaun Truelove[12], Jay K. Varma[26], Yonatan H. Grad[27]

**Affiliations:**

1. Center for Forecasting and Outbreak Analytics, US Centers for Disease Control and Prevention, Atlanta, GA, USA

2. Center for Communicable Disease Dynamics, Department of Epidemiology, Harvard T.H. Chan School of Public Health, Boston, MA, USA

3. François-Xavier Bagnoud Center for Health and Human Rights and Department of Social and Behavioral Sciences at the Harvard T.H. Chan School of Public Health, Boston, MA, USA

4. Boston Children's Hospital, Harvard Medical School, Boston, MA, USA

5. Department of Epidemiology and Public Health Medicine, Imperial College London, London, UK

6. Big Data Institute, Nuffield Department of Population Health, University of Oxford, Oxford, UK

7. Department of Epidemiology, Bloomberg School of Public Health, Johns Hopkins University, Baltimore, MD, USA

8. Division of Vector-Borne Diseases, US Centers for Disease Control and Prevention, San Juan, PR, USA

9. Department of Computer Science, University of Colorado Boulder, Boulder, CO, USA

10. BioFrontiers Institute, University of Colorado Boulder, Boulder, CO, USA

11. Office of Public Health Data, Surveillance, and Technology, US Centers for Disease Control and Prevention, Atlanta, GA, USA

12. Department of Epidemiology, UNC Gillings School of Public Health, Chapel Hill, NC, USA

13. Minnesota Department of Health, Minneapolis, MN, USA

14. Office of Advanced Molecular Detection, US Centers for Disease Control and Prevention, Atlanta, GA, USA

15. Massachusetts Department of Public Health, Boston, MA, USA

16. Department of Ecology and Evolutionary Biology, Princeton University, Princeton, NJ, USA

17. Department of Integrative Biology, University of Texas at Austin, Austin, TX, USA

18. Division of Disease Control, New York City Department of Health and Mental Hygiene, New York City, NY, USA

19. Pandemic Prevention Initiative, The Rockefeller Foundation, Washington DC, USA.

20. Department of Biostatistics, University of Massachusetts, Amherst, MA, USA

21. United Kingdom Health Security Agency, London, UK

22. Departments of Computer Science and Computational Biology, Carnegie Melon University, Pittsburgh, PA, USA

23. Division of Epidemiology, Department of Medicine, University of Utah, Salt Lake City, UT, USA

24. Siemens, San Diego, CA, USA

25. Division of Healthcare Quality Promotion, US Centers for Disease Control and Prevention, Atlanta, GA, USA

26. SIGA Technologies, New York City, NY, USA

27. Department of Immunology and Infectious Diseases, Harvard T.H. Chan School of Public Health, Boston, MA, USA

**Abstract**

The COVID-19 pandemic has highlighted the need to upgrade systems for infectious disease surveillance and forecasting and modeling of the spread of infection, both of which inform evidence-based public health guidance and policies. Here, we discuss requirements for an effective surveillance system to support decision making during a pandemic, drawing on the lessons of COVID-19 in the U.S., while looking to jurisdictions in the U.S. and beyond to learn lessons about the value of specific data types. In this report, we define the range of decisions for which surveillance data are required, the data elements needed to inform these decisions and to calibrate inputs and outputs of transmission-dynamic models, and the types of data needed to inform decisions by state, territorial, local, and tribal health authorities. We define actions needed to ensure that such data will be available and consider the contribution of such efforts to improving health equity.

**Introduction and Purpose**

To monitor pandemic pathogens effectively, modern surveillance systems should make use of the growing wealth of routine data from the health sector and beyond, and public health experts must integrate these data in new ways that increase their value. We need purpose-built systems to detect new and evolving threats and to provide information as quickly as possible about those threats. What are the characteristics of the new pathogens or new variants of existing pathogens? What is their incidence and prevalence? What is the vulnerability of the population to infection and disease? What is the impact of our efforts to respond to these threats?

Systems to generate, integrate, and interpret these data should be designed and built with the explicit purpose of providing timely evidence to inform decisions about disease control and mitigation. First, they will provide direct input into decision making. For example, evidence of low vaccine effectiveness may prompt efforts to boost or change formulations or doses. As another example, real-time lab order data for diagnostic tests may prompt adjustments to resource allocation. Second, these data will parameterize scenario and forecasting models (1, 2). For instance, estimates of per-case severity of a new variant, incorporated into forecasts or other models of case burden, may influence planning for hospital capacity and supply stockpiling and distribution.

This document reflects the framing ideas and the discussions held at a symposium organized by Harvard T.H. Chan School of Public Health entitled "Quantitative Tools and Data Opportunities for Pandemic Surveillance and Response," held June 29-30, 2022, involving a range of public health and public officials, surveillance experts and other epidemiologists, and epidemic modelers. We first aim to identify the most important decisions for disease control and mitigation and the evidence that is needed to inform them. We then describe a set of surveillance activities designed to provide timely, reliable, and appropriately scaled data to inform these decisions. Our focus in this report is limited to domestic detection, characterization, and estimation of the burden of a pandemic pathogen in terms of direct health effects. Although we note the importance of monitoring economic, social, and indirect public health impacts of a disease control measures, we do not offer a comprehensive treatment of this element of pandemic monitoring and response activities. This paper builds on earlier efforts (3) while incorporating both the new possibilities that technology now provides, as well as the lessons of COVID-19.

We differentiate between two related but distinct goals of surveillance, as this document will focus on only one. The first goal is to provide early warning about a potential pandemic, and so this type of surveillance includes global monitoring and rapid identification of domestic introductions. The second goal is to provide support for decision making during an ongoing pandemic, including tracking incidence, prevalence, and the pathogen's properties. While we will briefly remark on the former—surveillance for early warning—we will primarily focus on the latter—surveillance for decision making.

**Detecting a jurisdiction's first cases of a new disease**

The first set of decisions faced by a domestic public health jurisdiction, following the appearance of a pandemic threat somewhere else in the world, concerns the questions of whether, how, and to what extent to scale up a response to reduce the risk of importation or, if importation has happened, to control its spread within the jurisdiction. Measures to reduce importation via restriction or testing of inbound travelers may buy limited time to prepare (4, 5), though such measures lose relevance once local transmission is established (6).

To inform decisions about how to balance scarce public health resources between preventing importation vs. controlling local spread, it is critical to assess the risk that the infection has already arrived and started spreading within the jurisdiction. Testing and sequencing of specimens from international travelers at airports and analysis of wastewater from international flights may provide evidence of pathogen importation(7). Early evidence of local spread may come from informal communications among health care providers, reporting systems such as ProMED-mail (8), and "pre-health care" data (e.g., absenteeism, internet search queries). Signals may arise from monitoring of syndromes compatible with infections, or the volume, distribution, and results from clinical laboratory tests. They may also come from anomalous findings in sentinel and research efforts (e.g., the Seattle Flu Study at the start of COVID-19 (9)).

With each type of monitoring, there is first the question of what defines the signal we are looking for and then there is a tradeoff between having a highly sensitive and timely system capable of sounding an alarm early on one hand and producing too many false alarms on the other. In most situations, a high positive predictive value for such systems will be essential because the cost of responding to frequent false alarms is high. Much more work is needed to assess how to use and combine complementary monitoring signals to identify points at which an alarm should be escalated into a response.

**Surveillance for Decision Making During a Pandemic**

A comprehensive list of the decisions and guidance required in a pandemic would fill a much longer document than this. Still, based on the combined experience of the emergency response phase of COVID-19 (2020-2023) and H1N1 influenza (2009-10) pandemics, we propose a set of consistent themes that capture many of the major types of decisions arising both in the early days of each pandemic and throughout.

A brief, necessarily incomplete, list of such decisions follows, adapted from the list in (3):

1. Public health goals of a response (elimination, control, protection of high-risk groups, protection of health care functioning, or a combination) and overall scale of response needed to meet these goals.
2. Timing of scale-up and scale-down of response.
3. Choice of nonpharmaceutical countermeasures (individual-targeted such as quarantine, isolation, and personal protection; population-targeted such as closures). This includes decisions about the timing, magnitude, and geographic range of protective measures that may be socioeconomically costly. A related set of decisions concerns how such measures should be prioritized, i.e., who should receive protective equipment when it is scarce, and how closures should be targeted to reduce economic and social disruption.
4. Choice of medical countermeasures, including diagnostics, therapeutics, and vaccines. This includes decisions about development, stockpiling, procurement, expanding capacity (e.g., building alternative care sites), and more. Here too, questions of allocation and prioritization are central. This also includes planning for potential surges.
5. Specific policies for each of the issues above in special populations including vulnerable communities, and settings such as health care, schools, congregate settings, transport, etc.
6. Balance between community countermeasures to reduce severe disease or reduce transmission (e.g., allocation of resources to those at high risk of complications or high risk of transmission).
7. Design and implementation of staged alert systems to provide real-time risk awareness and trigger policy changes (10, 11).
8. Imposition and removal of international travel screening and restrictions.
9. Choice of public health communication strategies.

Each of these decisions requires specific data to decide how to improve health equitably, effectively, and efficiently while minimizing social and economic disruption. For example, decisions on testing, isolation, and quarantine policies require evidence on the natural history of infectiousness (or at least a proxy such as viral load), test sensitivity at different levels of viral shedding, the relationship between symptoms and infectiousness, and the potential economic and social consequences for various communities of the policies under consideration. In contrast, decisions about the timing of vaccine boosters require evidence on the effectiveness of existing vaccines against infection, transmission, and severe disease endpoints, stratified by such factors as pathogen variant, time since vaccination, and age, as well as understanding of how vaccine protection is distributed across demographic groups.

**Decisions faced by State, Territorial, Local, and Tribal Authorities**

In the federal system in the U.S., public health is decentralized and typically not coordinated among states. State, territorial, local, and tribal (STLT) governments are responsible for nearly all binding policy decisions in public health, with governance health structures varying by state (12). The purview of these bodies includes (13) prescribing and enforcing isolation, quarantine, mask mandates, and restrictions on businesses and gatherings; vaccine prioritization and distribution; and (to a degree) diagnostic testing. They also hold responsibility for closely related areas, such as public education. STLT governments all have a desire for similar types of data, but vary in how much they need, how quickly they need it, and how they use it.

Many decisions involve procurement and distribution of countermeasures. Because STLT authorities are making allocation decisions within their jurisdictions (e.g., for counties, cities, hospitals, schools), jurisdiction-wide measures of disease activity are rarely sufficient; instead, more geographically granular numbers are required.

Table 1 summarizes key decisions and associated needs for jurisdiction-level data and analytics in COVID-19 cited by state and local leaders during the symposium:

| Decision | Data/analytics need |
|---|---|
| Size of response needed | Rapid threat characterization |
| Choice of community countermeasures | County-level disease burden and transmission measures |
| How to ensure adequate supply of hospital beds, ventilators, personal protective equipment | Forecasts of demand for these items |
| School and congregate setting policies (closure of schools, infection control measures in jails, prisons, nursing homes, etc.) | Understanding of rates of transmission into, within, and from each of these settings and impact of testing and infection control on these rates as well as population specific health-risks |
| Countermeasure deployment within a jurisdiction | Age, racial, ethnic, and geographic patterns of transmission and disease burden. Note: these are often crude proxies for social determinants of infection and outcome risk, not adequate for scientific understanding of why particular groups are at risk, but nonetheless potentially useful for focusing prevention and treatment efforts on those with high vulnerability. |

| Efforts to distribute and promote vaccination | Variant prevalence, vaccine coverage, and vaccine effectiveness against dominant and emerging variants |

**Data needs for decision support: the COVID-19 experience**

A range of data sources could and, during the COVID-19 pandemic, did provide evidence to support decisions by health authorities. Following initial social media reports of clusters of pneumonia, some of the earliest specific data to characterize the COVID-19 threat came from traditional sources, such as from case reports posted on Chinese public health department websites (14). A key challenge was the repeated change in the syndromic case definition in the early days (15). But other early data came from unexpected sources, such as cruise ships (16), restaurants (17), and fishing vessels (18), where conditions allowed inference of the path of transmission and thereby provided evidence about the degree and mechanisms of spread. Specifically, these provided some of the earliest evidence of asymptomatic and aerosol spread, which, when properly interpreted, aided in the design and prioritization of testing and other control measures. As had been true in the 2009 influenza pandemic (19, 20), sampling of travelers provided early estimates of the extent of global spread, growth rates, and likely under-detection (21, 22).

As the pandemic spread, the strengths and limitations of each data source became evident. Multiple data types were required to provide even an incomplete picture of trends in incidence and prevalence and behavior (23, 24). For example, case counts were used as an important indicator of disease burden. However, the relationship between new cases and true incidence varies as a function of numerous factors, including test availability, test reporting requirements by jurisdiction (which did not always include reports of negative tests), rates of testing through clinical facilities (which declined with the growth of rapid antigen testing), and incentives to get or avoid testing (Fig 1). Some of these limitations can be mitigated by breaking out case counts by the reason why an individual was tested (symptoms, travel, surveillance), but this was not consistently done in the US. As a result of these limitations, hospitalizations and even deaths were increasingly used as the more reliable indicators of case numbers, sacrificing some timeliness for a more consistent relationship to the underlying incidence of infection. Random sampling approaches (described below) can overcome these limitations and provide more consistent and reliable estimates of incidence and prevalence and how these change over time. Only the UK and Luxembourg used random sampling on a large scale, perhaps because of the cost and logistical challenges. Notwithstanding their limitations, case counts were the major early data source in the U.S. and provided critical evidence especially when linked to demographic information. Syndromic surveillance—done routinely as part of monitoring influenza trends—from emergency room visits and hospital admissions were also valuable data sources, particularly when testing was limited. However, interpretation of syndromic surveillance was complicated by changes in healthcare seeking behavior and the increased use of telemedicine.

Novel data streams provided confirmatory evidence as well as early warnings of trends that might not be evident in case counts. For example, wastewater surveillance for SARS-CoV-2 was adopted in numerous jurisdictions from 2020 to 2022 and provided evidence on local epidemic trends, although the precise relationship between wastewater abundance and the number of infected persons depends on the wastewater sampling scheme and on shedding patterns (among other issues), and thus difficult to quantify (25). Moreover, by its nature, wastewater cannot indicate who has been infected, thus leaving the demographic profile of infected persons uncertain. Finally, wastewater surveillance as currently applied will miss infections in areas with high reliance on septic systems, which serve roughly one fifth of the households in the U.S. with heavy concentrations in certain geographies, inducing inequities in whose infections are tracked by this approach (26).

Another novel data stream is the use of population-wide distributions of viral load measured from PCR testing, which, in the aggregate, can provide information on the trajectory of the epidemic, even from a single cross-sectional analysis (27). This approach has reached proof of concept and has the advantage that it may be less sensitive to trends in testing behaviors than measures of incidence based on case counts, and unlike wastewater surveillance, it can provide some information on the demography and precise location of cases. However, further work is needed to see how a transition to non-PCR testing for many new cases, the halting of pre-procedural and asymptomatic testing, and the shifts in viral kinetics that come with immunity (from vaccination and infection) affects the nature of this signal (28-30) . Moreover, the identifiability of time-since-infection from viral load, which is needed for the approach to work, depends on the asymmetry in viral load over time (fast rise, slower decline (27)), which may or may not be a feature of future infectious diseases.

Digital data can also be used for surveillance and to inform on epidemic trajectory. ProMED-mail (8) and HealthMap (31) are valuable for flagging and disseminating reporting and information on events known or suspected to be infections and outbreaks [refs]. Data from search engines, social media, and news reports data can also inform epidemic dynamics and for forecasting (32-34).

Finally, testing for antibodies in sera collected either for the purpose of serologic surveillance or in convenience samples (e.g., blood banks, discarded clinical samples) was used to characterize both the landscape of population immunity (i.e., who was and wasn't vulnerable to reinfection) and to distinguish between those who had acquired immunity via vaccination vs. infection (35). Secondary analyses of COVID-19 vaccine studies identified complexities in answering the latter question, finding that infection does not reliably induce antibodies to non-vaccine antigens in vaccinated individuals (36).

An important conclusion is that no one data source or surveillance tactic is sufficient. In a setting like the U.S., multiple surveillance approaches are needed at scale. Beyond the obvious need to combine data sources, several points stand out.

The first is the value of data completeness and of linking data types to produce evidence that is greater than the sum of the parts. For instance, while counts of cases and hospitalizations are valuable, missing race/ethnicity, geographic, and other patient characteristics have impeded efforts to improve services to groups that are underserved or experience high disease burden and to improve equity in health-related outcomes. Similarly, meticulously linking sequence data from patient isolates with demographic and clinical predictors of severe outcomes, including vaccination history, and clinical outcomes can help to evaluate the threat posed by novel variants (37). Unfortunately, despite prodigious amounts of SARS-CoV-2 sequencing in the U.S., this form of linkage has been relatively rare to date.

Second is the value of clear and accessible data dashboards with transparent data sources to make the state of the epidemic locally evident to the public. The same data should also be available to analysts in public health departments, academia, and other sectors via application programming interfaces (APIs) to facilitate rapid data analysis. This can facilitate shared decision making and help to increase public support for control measures. For example, the city of Austin, Texas developed a COVID-19 staged alert system that guided local policy between May 2020 and March 2022 (10, 11). The public-facing dashboard featured a single graph that tracked COVID-19 hospital admissions and clearly indicated thresholds between the red, orange, yellow, green, and blue risk levels that were linked to specific actions. The county judge, city mayor, and public health authority cited the dashboard almost daily to communicate risks, explain changes in policy, and cultivate adherence via news outlets and social media.

This system was only possible because local authorities required area hospitals to report daily admissions beginning in April 2020, long before such data were generally available.

Finally, discussion at the Symposium emphasized the value of metrics that could be compared across jurisdictions. Decision-makers expressed a desire for objective criteria by which their performance can be judged. Comparisons across states, for example, were hampered by differential testing rates that affected case counts in ways not reflecting actual prevalence. A CDC-supported academic effort called covidestim (38) used Bayesian evidence synthesis to harmonize estimates of current and cumulative infections across states and counties, providing an example of what could be done by health authorities. However, this effort was also hampered by unanticipated changes in reporting tempo, as well as 'data dumps' and data backfilling. Different definitions of COVID-19 hospitalization across states and over time impeded comparisons of outcomes that would have provided indications to elected leaders of the quality of their responses and informed improved responses.

**Surveillance inputs to Forecasts, Scenario Projections, and Analytic Models**

As noted above, many aspects of pandemic decision-making can directly incorporate evidence from surveillance, and will also make use of

- Nowcasts: estimates of current burden of cases, hospitalizations, deaths, and other quantities that account for delays in reporting;
- Forecasts: relatively short-term projections of case, morbidity, and mortality burden, typically on the scale of days to weeks;
- Scenario models: longer-term estimates of pandemic dynamics that anticipate multiple possible futures under stated assumptions about behavior, viral evolution, vaccine durability, etc., typically on the scale of months to years;
- Results from analytic models: estimates about different characteristics of the pathogen or a population of concern that are specifically designed to inform a decision or guidance, such as school- or nursing-home-based testing policies, contact tracing procedures, or quarantine duration.

These categories of decision-support tools require estimates of *input quantities* that represent the assumptions of the models—for example, in scenario models, estimates of per-case severity or vaccine effectiveness. These estimates need to be timely, representative, and specific to the pathogen or variants circulating or anticipated to circulate (for example, due to importation). In addition, scenario and forecast model *output* must be calibrated to existing measurements of disease burden: incidence of infection, diagnosed cases, hospitalizations, deaths, and other relevant metrics, as well as against cumulative measures such as seroprevalence. The categories of input and output are somewhat fluid, as a model with sufficient data to calibrate outputs may be able to estimate the values of some of the quantities described here as inputs. In a fully Bayesian framework, both external estimates (as priors) and calibration to output data may contribute to posterior parameter estimates.

For forecasts, evaluation can be performed quickly due to the short-term horizon of the predictions made, with results that can provide feedback to modelers about places where models are mis-specified. Evaluating scenario projections is more complicated, as multiple sets of counterfactual projections are made under different assumptions about how a pandemic situation will evolve over the course of months or years(39). Most (or perhaps all) of the scenarios will not be realized exactly as assumed, making evaluation less straightforward.

Together the quality and timeliness of these input parameters and output calibrations are important determinants of how useful a model is for decision making. While there are techniques to adjust for incomplete or lagged information, the absence of certain ingredients—especially model output calibration targets such as numbers of cases or hospitalizations—can critically compromise the ability to generate models that reflect reality to the point of hampering basic situational awareness. Data systems that support modeling and in turn decision-making during pandemics should be considered vital national security capabilities and prioritized accordingly.

A list of the key needs for model inputs is as follows, many of which may change as a pathogen evolves (referred to below by their letters):

a. Pathogen kinetics/epidemiological parameters (e.g., incubation period, latent period, infectious period, infection fatality ratio). Estimation of these inputs may itself require simple models, particularly at the early stages of a pandemic [ref: Gostic paper].
b. Transmissibility and efficiency of various transmission mechanisms
c. Risk factors for infection and severity
d. Individual and population immunity (including effects of infection, vaccination, and waning)
e. Diagnostic test characteristics, including specificity and sensitivity for active (acute) and past infection
f. Vaccine effectiveness and waning of effectiveness, for infection, severe disease, and mortality endpoints.
g. Treatment effectiveness
h. Policies, uptake, and effectiveness of nonpharmaceutical interventions
i. Population mobility and interactions: contact networks and patterns by setting
j. Importation risk
k. Other co-circulating pathogens of concern (e.g., if concurrent with significant influenza transmission)
l. Capacity and utilization of healthcare resources (including hospital beds, therapeutics, and vaccines)

Key additional data requirements for fitting models– as well as for general situational awareness – include:

m. Geographically and demographically stratified incidence, duration and prevalence of infection, hospitalization, ICU admission, death, and other relevant metrics associated with the pathogen, ideally by variant.
n. Strain-specific incidence.

**Meeting these needs**

This draft framework is a preliminary attempt to scope a system that could meet the needs listed above for situational awareness, decision support, and inputs and outputs for modeling and analytics for a new variant or a new pandemic. Capacity to achieve these would also be applicable to other pathogens, especially, but not only, respiratory ones.

**A. Estimating model inputs**

System 1: High-frequency sampling for pathogen kinetics and diagnostic sensitivity (quantities a,e)

Possible Mechanism: Surveillance would be established to obtain repeated samples (for COVID-19, respiratory samples) from individuals exposed to a pathogen of interest (now, SARS-CoV-2) from the time of exposure through infection to the time of clearance. High-frequency sampling will provide detailed profiles of pathogen kinetics, which could be subgrouped by prior infection history, vaccination status, pathogen variant, demographics, and other predictors. Simultaneous use of nucleic acid amplification (NAAT), culture, and antigen-based testing on these specimens would provide detailed estimates of the sensitivity of each as a function of symptoms, pathogen load, pathogen infectious capacity, variant, and time since exposure/first-positive to inform choice of diagnostics and isolation/test/quarantine policy.

Performers might be research/surveillance networks or STLT health departments (recognizing that the health departments may have limited bandwidth in the context of an outbreak). The ability to scale up is critical. While pathogen kinetics are not likely to vary from place to place, geographic diversity in sites capable of performing these investigations will increase the timeliness of results in case one region is hit much earlier than others.

Settings may include households, universities, day cares and schools; intensely monitored cohorts such as sports leagues or health care workers, congregate settings such as homeless shelters, correctional and detention facilities, or nursing homes.

Precedents: UK Household study (40) and U.S. National Basketball Association studies (29, 41).

System 2: Integrating routine sequencing with detailed clinical data (quantities b,c,d,f,g,n)

Possible Mechanism: A payer-provider network with diverse geographic and demographic representation (alternatively, a private sector entity or consortium of public health departments and laboratories capable of merging clinical data with sequence data) would track individuals as a cohort (not necessarily defined by long-term follow-up but perhaps with exposure or a positive test as an entry criterion) with known vaccine and prior infection history through diagnosis (outpatient or inpatient) and through the cascade of care to estimate the probability and severity of infection as a function of this history (vaccine effectiveness and infection-acquired immunity) and variant. Sequencing of positive clinical specimens would enable the variant-specific estimates. This system would provide a reliable infrastructure for assessing severity, vaccine effectiveness, and treatment effectiveness linked to infection and vaccination history for each new variant/virus. It would be crucial to link electronic health record (EHR) within the network to key external sources of data such as immunization registries. Improving completeness of such registries is also a high priority to improve the quality of these inferences. Strategies for linking pathogen genome sequencing with EHRs will depend on whether these data are from clinically validated systems and, if not, will require consideration to ensure use for research and not clinical purposes.

It would be valuable to explore to what extent such studies could be done in networks such as PCORNet (42) or the Vaccine Safety Datalink (43) that assemble EHRs from multiple health systems into a common data model; questions include how rapidly this could be done and whether sequence data could be linked to these records.

In addition to payer-providers, robust testing, reporting, and data collection capabilities should be considered for congregate settings at high-risk for transmission such as skilled nursing facilities, correctional facilities, detention facilities, and homeless shelters that can follow individuals from positive tests through outcomes.

Precedents: Cohort studies on variant-specific relative severity (44), relative vaccine effectiveness (45) and absolute vaccine effectiveness (46) have been performed during the COVID-19 pandemic. None of these included genomic sequencing or serological profiling integrated with clinical data collection, in part due to the issues of linking with EHRs as mentioned above. Integration of sequencing in particular is essential for the likely future scenario where one cannot rely on proxies for genetic variant that have been exceptionally convenient in COVID-19, notably the failure of the S-gene PCR target in certain polymerase chain reaction-based diagnostic tests.

In the US, this work could build upon or integrate with existing platforms such as VISION and Investigating Respiratory Viruses in the Acutely Ill (IVY) (47). Key additions would be sequencing and more comprehensive estimates of severity.

System 3: Behavioral surveillance and other routine data collection (quantities h,i,j,l).

Goals of behavioral surveillance are to provide real-time estimates of mobility, work-from-home frequency, proportion of schools open or closed, use of other nonpharmaceutical interventions such as masks, and vaccine behavior/hesitancy. Data useful during COVID-19 included vaccine coverage from HHS Protect (48)and Census Pulse (49) and other surveys on vaccine intentions, mask use, work-from-home, and school opening/closure. Private sector (e.g., mobility(50)) and publicly available data (e.g., (51) exist that measure many quantities of interest. These include self-reported mask use, absenteeism data from school and work, internet search queries, and much more. Further work needs to be done in several areas to enhance the value of these data streams:

- identify cost-effective sources of such data;
- quantify the degrees of representativeness in measurement from these different data sources by such factors geography, race/ethnicity, and social determinants of health;
- improve our mechanistic understanding of how these measures of mobility relate to transmission behavior, which will likely differ by social factors, pathogen transmission routes, and epidemic stage, among other factors (24)

A particular example of one such data stream is air travel and other travel data to estimate importation risk.

Precedent: Census Pulse and other surveys exist. Many local jurisdictions have used mobility data from private providers, often via academic intermediaries (e.g., https://www.covid19mobility.org/) to assess local trends. Vaccine coverage data exist with some limitations. The Center for Disease Control and Prevention's (CDC) Division of Global Migration and Quarantine maintains access to timely estimates of air travel volume.

**B. Fitting model outputs**

System 4: Repeated testing for infection and immunity in a random sample of the population (m)

Mechanism: An academic, government (e.g., CDC or a coalition of state health departments), or private sector entity would identify a longitudinal sample and/or repeated cross-sections representative of the U.S. population for monthly testing for infection and immunity as evidence of prior infection. In the COVID-19 case, this would be PCR testing of respiratory samples and antibody measurement in blood; testing approaches might differ for future pathogens. Samples would be obtained by home visit or mail/courier. Specimens testing positive for one or more respiratory viruses would be sequenced. The initial sample would be powered to detect US-level trends; scale-up in a pandemic would enable regional/state-level and demographic-specific (e.g., age, race, sex-specific) estimates of virus prevalence and seroprevalence irrespective of symptoms and at the level of variant/subtype/species/type (depending on the pathogen).

In pathogens with antibody-based immunity, blood samples would be tested for multiple antibodies including vaccine and nonvaccine antigens of the novel pathogen. These would provide a population-based denominator for severity estimates, enable calibration of scenario and forecast models, track trends in viral species/variants in an unbiased way, and estimate the magnitudes of health inequities to better prioritize prevention measures (52).

Addition of serologic testing of a random, representative sample of the population would supplement existing passive serosurveillance approaches such as from blood donors (35), newborn heel sticks (53), or discarded specimens(54, 55). Longitudinal sampling would enable more precise estimates of rates of waning of antibody concentrations (56, 57) and the consequences for estimation of cumulative incidence using particular assays.

An alternative approach would be to use healthcare-based testing of individuals requiring admission for conditions not directly related to the pandemic, using weighting to standardize the population seeking health care to the background population (58, 59), though the quality of such data would need continuing validation.

Other alternatives would include the use of voluntary testing results, such as those gathered by test-proctoring telehealth services, retail pharmacies, or the like. CDC/FDA requirements to ask the reason for a test would facilitate interpretation (symptomatic vs. travel vs. exposure, for example).

Precedents: The main proposal could be roughly modeled on the United Kingdom COVID-19 Infection Survey and REACT-1 studies. One of the alternative approaches—universal testing of individuals requiring admission for non-pandemic reasons—was used in New York City early in the COVID-19 pandemic (60) and has been used in Indiana with reported high value (58, 59) for both prevalence and seroprevalence.

System 5: Maintain hospitalization surveillance data (l,m)

Hospitals have been required to report COVID-19 and influenza hospitalizations to HHS, and these formed the backbone of multiple forecasting and scenario modeling efforts in the US. It is critical to maintain the generation, interpretation, timeliness, and accuracy of these data to inform forecasts. In addition to the forecasting products, these data underlie hospital capacity and burden situational awareness, the ability to monitor outbreaks, and community burden indicators.

Precedents: Exists as of September 2023 but needs to be maintained at a base level outside of emergencies and be able to ramp up quickly at a time of new emergency (61).

**C. Actions needed**

***Administrative and reporting preparedness***
The response to COVID-19 required collaborations across sectors—public, private, and academic—but these collaborations were often forced to work through administrative frameworks that were not designed with speed and flexibility in mind. In turn, such mis-specified frameworks ultimately slowed or limited some critical public health projects and prevented others from being undertaken entirely. To address this class of problem, we propose six ideas below that would update, recast, or create key frameworks that establish links across sectors and that facilitate the urgent work of pandemics, while maintaining safeguards and oversight.

1. *Emergency data use agreements and formats.* Data use agreements (DUAs) are core elements to collaborative work across institutions, but they pose two types of challenges. First, the process for negotiating an agreement acceptable to the institutions providing and receiving the data is often slow. The staff on each side tasked with reviewing and signing off on these agreements may have many competing priorities or be overwhelmed as an outbreak or pandemic may dramatically increase the volume of DUAs. Work on a sensitive or high-profile project, such as associated with an outbreak of infectious disease or a pandemic, generates additional scrutiny and often further lengthens the review process. Second, conflicting limitations can stall progress or even undermine a project before it starts. For example, in a partnership between academics and government public health institutions, academic institutions may deem the freedom to publish without interference to be non-negotiable. Public health institutions, however, may require veto power over what, if anything, is published, due to the sensitivity of the institution's data and ownership thereof. To address these problems, one solution is to establish Emergency Use Data Authorizations (EUDAs) for public health data with a standing framework vetted and updated regularly (e.g., annually), perhaps at the individual state level. Such EUDAs would catalyze collaborations and enable investigators at both institutions to shift the balance of effort up front from administrative to research tasks. As these are put in place, discussions about data formats can take place, ideally also in advance, to ensure that when data are delivered they are as ready-to-use as possible.
2. *Surveillance versus research: updating the Common Rule*. Projects designated as human subjects research require institutional review board (IRB) review, whereas those designated as public health surveillance are deemed not to be research, and thus do not require IRB review. This surveillance-research dichotomy has substantial implications for timeliness and speed of work, because writing, reviewing, and adjudicating IRB reviews—while vitally important for protecting the rights, welfare, and well-being of human subjects—may take days to weeks. The boundaries between surveillance and research are governed by the Common Rule, which states that public health surveillance activities "include those associated with providing timely situational awareness and priority setting during the course of an event or crisis that threatens public health (including natural or man-made disasters)" (45 CFR 46.102(*l*)(2)) (62). Unfortunately, these boundaries lacked clarity and standardization as questions arose during the COVID-19 pandemic. For example, while case monitoring is clearly surveillance and a routine public health activity, one could make a strong argument that "situational awareness and priority setting" includes assessing vaccine effectiveness and disease severity for new variants. However, analysis of variants requires pathogen genome sequencing, which is viewed by some as constituting research, as is evaluation of vaccine effectiveness, another critical public health function which is not exclusively a research objective. Modifying the text of the Common Rule to explicitly include examples such as these or providing an interpretation of the surveillance/situational awareness exemption that includes these activities would considerably improve the ability for public health agencies to maintain situational awareness and set priorities, quite in line with the spirit of the exemption.
3. *Streamlined IRBs.* Where projects fall under human subjects research designation and require IRB review, generic, pathogen-agnostic study protocols for specific populations would accelerate research by decreasing the time to first data. Preapproval of a range of well-defined studies targeted at emergency response and using specific data sets would retain critical protections for human subjects, while allowing high-urgency protocols to be "on the shelf" and ready for fast rollout. As an additional feature, such preapproved protocols would also free up valuable researcher and IRB reviewer time, having converted per-submission efforts during a pandemic into fixed-cost efforts ahead of time. Moreover, designing consent processes for normal "peacetime" studies to allow use of data and specimens in public health emergencies could avoid some of the delays experienced during COVID-19 with, for example, use of the Seattle Flu Study's specimens to understand early transmission of the virus in the U.S. (9).
4. *Case reporting standardization*. Tracking and understanding outbreaks, particularly at their beginnings, rely on case reporting. Ideally, public health efforts would follow case trends over time and across regions, compare and monitor clinical features including disease progression, resolution, and response to interventions, and track demographics of infected individuals. But lack of standardization of case report protocols, parallel or overlapping surveillance systems that result in duplication (often with varyingly completed fields for the same case), and inadequate systems for incorporating updates as further information about a case accumulates after the initial report, among other issues, result in case report data that require much time and effort to sort through. Worse, these issues may render some fraction of case reports unreliable. Improving national surveillance systems to be more uniform, timely, and flexible could serve both local and national surveillance needs would help address these issues (60).
5. *Dataset accessibility*. In the absence of a U.S. national healthcare system, research into the distribution and burden of clinical conditions depends on academic or private data streams, including surveys and surveillance systems constructed to address specific questions, and databases of insurance claims which represent utilization of the healthcare system. Insurance claims datasets include those from (i) employer-based insurance companies (e.g., MarketScan ) (ii) all-payer claims databases available in some states (which, since a 2016 Supreme Court decision (63), are no longer necessarily 'all payers') , (iii) Medicare for individuals over 65 years of age, (iv) Medicaid, which provides coverage to over 18% of the US; and (v) data bases for other specific populations, such as those of the Veterans Affairs Health System, the Indian Health Service, and the Department of Defense. While these datasets can provide an important window into healthcare use across demographics and geography, access to these datasets can be expensive and time and labor intensive. Gaining access to Medicaid data, for example, presents a substantial burden, since this has to be acquired on a state-by-state basis. Establishing standing flexible DUAs for these datasets, with a single agreement across states for Medicaid and other state-controlled data, could enable both routine surveillance-type analysis to identify trends (such as disease outbreaks or patterns of disease spread) and to evaluate the impact of clinical and public health interventions.
6. *Public health-health care partnerships*: While the U.S. does not have a national health system for all, it has a wealth of data in the health care sector that can inform public health decision making. Multiple studies at the Centers for Disease Control and Prevention (CDC) and other institutions harnessed such data to provide estimates of key quantities such as vaccine effectiveness (64, 65), variant severity (66), and antiviral effectiveness (67), as well as for surveillance of disease burden and its correlates (68). Building public health partnerships with the health care sector in advance to set in place the administrative, information technology, and financial arrangements to make possible high-quality analyses of this sort rapidly (and automated where possible) would greatly increase the timeliness and value of such efforts (69).

***Strengthening personnel and research ties***
In response to the COVID-19 emergency and the need for expertise to gather, analyze, and interpret evidence around the pandemic and the clinical and public health responses, many academics put aside their usual research programs to engage directly in public health activities and research. The close interactions between academics and local, state, and national public health officials were often productive and important for guiding the pandemic response but raised issues that should be addressed before the next pandemic. These include the ad hoc way in which these academic-public health collaborations came into being, the lack of uniformity of access to academics with appropriate expertise across states, and the misalignment of incentives between public health and academic work.

Ideally, academia-public health collaborations can be rapidly scaled up in times of need through established pathways. One idea is to create a "rotator" program, in which academics (and potentially those in training, including doctoral students and postdoctoral fellows) are embedded within public health agencies—and similarly public health officials are embedded within academic groups—for intervals (such as 3 or 6 months) that build familiarity, collegiality, and accessibility. The LEAP fellowship through the Infectious Disease Society of America (70) and the joint Infectious Diseases / EIS fellowship (71)programs are efforts in this direction. Another approach is to establish an academic career path in which some fraction of time and effort are based in public health activities, analogous to academic medicine paths in which researchers spend some fraction of their time doing clinical work. Cooperative agreements established in 2023 between the CDC's Center for Forecasting and Outbreak Analytics (CFA) and academic and other groups include a surge provision whereby the performers on these agreements would provide scientific assistance in times of crisis. Relatedly, an official "public health reserve corps" of analysts and modelers could provide a workforce available to be called up to prepare for and respond to emergencies. Formal recognition of these paths as prestigious and vital, and placing value on these activities within the academic systems of rewards and incentives, will be key to success.

Often tools developed for one public health jurisdiction solve common problems and could be readily implemented in other jurisdictions, underscoring the importance of making code open access and ideally making tools generalizable. This would have the benefits of "not reinventing the wheel" and allowing those jurisdictions lacking local expertise access to useful tools. A curated clearinghouse of such tools, organized by research problem and perhaps hosted by CFA offers one strategy for providing access. Broad efforts to create and maintain state-of-the-art tools for epidemiologic modeling, such as Epiverse (72) and Recon are encouraging developments in this space (73).

*Educating the consumers*
Tools are best deployed by those who understand how they work, how they are limited, and how they can be modified to improve their applicability to specific situations.  Workforce development mandates to build subject matter expertise within public health departments, such as through CDC efforts via the Office of Science and the Office of Advanced Molecular Detection and through fellowships such as the CDC/Association of Public Health Laboratories (APHL) program, are critical efforts. While waiting for these workforce programs to get up and running, and since public health officials may not stay current with the frontiers of analytical and modeling methods, opportunities for regular formal trainings should be developed. For example, meetings such as the Council of State and Territorial Epidemiologists Annual Conference could provide a forum for workshops on advances in modeling, genomic epidemiology, and other fields. Relatedly, encouraging public health officials to attend field-specific meetings (e.g., Epidemics, Applied Bioinformatics in Public Health Microbiology) could provide opportunities for knowledge sharing, relationship building, and networking across sectors and disciplines.

*Improving knowledge flow*
Successful communication of a health agency's current understanding of a pandemic and outlook for its future requires a combination of approaches to communicate different kinds of data and outlooks, for different audiences. It has been suggested that principles for such communications include: thematic structure related to informing key decisions, synthesis of evidence from multiple sources, quantification of uncertainty, inclusion of visualizations as well as text and tables, and inclusion of forward-looking material (outlooks for the future) (74); another important principle is open access to the data underlying figures in these reports. CDC's Technical Reports on the Mpox epidemic in 2022 (75) sought to put these principles into practice, explicitly emulating aspects of the UK Health Security Agency (UKHSA) Technical Briefings from COVID-19 (76). Creating a regular cadence for such reports during an emergency, as was the case in the UK during the height of the COVID-19 pandemic, may help develop an audience and facilitate knowledge flow.

*Centering equity*
The World Health Organization has stated that "Countries have an obligation to develop appropriate, feasible, sustainable public health surveillance systems" to ensure that the health needs of populations are quantified so that they can be addressed. While there has been a disproportionate impact of COVID-19 on racial and ethnic minorities and on socioeconomically disadvantaged populations in the U.S. (77) and elsewhere (78, 79), a persistent problem is that race/ethnicity data are too often missing from surveillance data. Under the plausible hypothesis that those with missing data on race/ethnicity are among the most disadvantaged, these missing data could lead to attenuated estimates of the degree of inequities; whether or not this is the case, it reduces the quality of the estimates by adding uncertainty. Improving the completeness of race-ethnicity reporting is an urgent priority to maximize the value of surveillance data to enhance health equity. Some symposium participants, while agreeing with the need for better reporting of such data, argued that in the presence of ongoing racial segregation, ZIP code or other geographic tags can be a useful proxy when such data are unavailable. Early maps of COVID-19 in New York City showed a higher prevalence of COVID-19 diagnoses in areas that were home to largely Black and Hispanic populations, as well as some areas where most residents were White and many believed to be first-responders (60). This was reflected in elevated COVID-19 mortality rates among Black, Hispanic, and Native American populations compared to White populations throughout the U.S., particularly in the early waves of the pandemic (80).

The Presidential COVID-19 Health Equity Task Force final report from 2021 (81) recommends strategies for enhancing equity in data, analytics, and research. These recommendations include standardizing demographic and socioeconomic categories, supporting equity-centered data collection, tracking and reporting on health outcomes for people in congregate and high-risk settings, and research and analysis on behavioral health. In a similar spirit, for any clinical or public health intervention, one should ask in what ways the intervention exacerbates or alleviates inequities. To put this into practice, one goal is the development of real-time metrics that inform municipalities and states on the equity of interventions and health outcomes, enabling adjustments and responses to keep equity at the forefront of intervention decisions.

Other important examples of the links between surveillance and health equity were discussed during the symposium. While documenting disparate impacts is a necessary starting point, identifying appropriate measures to rectify these inequities will often require understanding where, why, when, and how they arise (82). An early example in the U.S. was a documentation that higher SARS-CoV-2 prevalence among mothers admitted for labor and delivery in New York City was associated with residence in boroughs with smaller reductions in mobility, suggesting inability to work from home as a potential driver of risk (60). Subsequent modeling work explained racial disparities in infection rates in U.S. cities as a consequence of higher exposure by minority groups not only to infection generally, but particularly to more crowded venues with higher infection risk (50). The age distribution of mortality by race/ethnicity (83), with its skew to younger ages among Black, Hispanic, and Indigenous individuals, also pointed to an increased exposure risk. The UK Government's Race Disparity Unit published a series of reports through the first two years of the pandemic enumerating hypotheses for mechanisms to explain disparate impacts, stating the current evidence related to these hypotheses, and recommending actions to address these drivers of higher incidence and severity in racial and ethnic minorities (52). In the U.S., such studies may require linkage of disparate data bases to identify where disparities arise during the cascade of care (84), a strategy that has long been useful in HIV/AIDS surveillance to understand loss points in the continuum of care (85, 86). For COVID, a full cascade would require an estimate of the actual number of infected individuals, the number of people who have been identified by testing (reflecting under-diagnosis), the number treated when treatment became available (reflecting under-treatment), the number hospitalized (reflecting access to care and disease severity), and the number of fatalities, jointly stratified by race and ethnicity, age group and sex. Ascertaining these would require both modeling-based estimates and data from multiple sources (e.g., clinical laboratories and vital registries). For example, an analysis from a New York City hospital suggested no racial difference in case fatality among hospitalized patients, supporting the idea that racial differences in exposure (more infections) rather than racial differences in outcome contributed to racial differences in overall mortality (87).

Other sources of inequity can affect case ascertainment and thus identification of opportunities for intervention. Geographic and temporal variation in testing effort in the U.S. was very large, resulting in difficulties in comparing incidence across jurisdictions. Rural areas were often the least able to access testing, though there were important exceptions (88). Notably, the use of random sampling stratified by geography mitigated this problem significantly in the UK (89), though it did not solve it entirely because participation was of necessity voluntary. Equity considerations may change as public health authorities rely on new data sources; for example, mobility estimates may depend on smartphone ownership, while wastewater surveillance for pathogen abundance will be unavailable in areas using septic systems (26, 90).

*Expanding the range of data types*
As described above, any health system, but particularly one as decentralized as that of the U.S., benefits from the ability to ingest and synthesize multiple types of data. Increased use of wastewater data (91, 92) has contributed to early warning of rising infection incidence and to surveillance for new variants. Further work to standardize collection and better define the quantitative relationships between true infection incidence and total and variant-specific concentrations of viral genomes in wastewater is needed to improve the value of such data, as well as a clear mapping of where it will not be informative, such as areas using septic systems. Likewise, mobility data from various sources (50) can be useful in informing strategies for disease monitoring and surveillance, modeling disease spread, and guiding interventions. Immune measures from serology provide a window onto past infection and a lens onto the landscape of risk (93). Here, further work is needed to ensure data standardization and accuracy as well as routine and frequent updating to capture important temporal variations. Such new forms of data may also raise privacy considerations that have not entirely been solved (94, 95).

Crowdsourced and survey data (96, 97) can provide important insights into behaviors that affect the interpretations of other data; for example, the increasing prevalence of self-testing using antigen tests for COVID-19 reduces the utility of PCR-positive case counts.

A key to making use of this expanded range of data types is solving the problem of how to synthesize multiple data types into a single estimate of a quantity of interest, accounting for the different properties of each data type (68), including understanding the different biases that will affect each data stream. Significant further work is needed to advance the ability to do this in real time.  A related but distinct problem is how to link data across data systems to understand the continuum of care and otherwise improve inference about the course of individual cases.

*Expanding the range of data sources*
The use of claims data from health care payers (insurers) and electronic medical records from providers has exploded in many areas of health services research. There have been some notable examples of such data for surveillance to address the questions described in this report (45, 64-66), but in the U.S. there remains untapped potential to expand such efforts and improve their timeliness. This will require building relationships between public health entities and health care systems in their jurisdictions, including relationships between scientific investigators in each sector with regular discussions for bidirectional learning. In the spirit of administrative preparedness above, this will require up-front planning of master agreements to move resources in a timely fashion to address pressing questions. Health providers and public health have suffered from a "two cultures" challenge that results in the need to expand public health training of investigators and other personnel in health systems, acknowledge the contributions of health systems to community benefits, and find ways to produce incentives so that contributing to public health surveillance aligns with the business interests of health systems. Medical examiners and coroners are another group that has been disconnected from public health but with whom cooperation can enhance and help to calibrate surveillance for pathogen-specific deaths, as illustrated by some examples both domestically(98) and abroad (99).

As we described above, new data sources become useful in proportion to our understanding of their "normal" behavior. As we expand the range of data types, it will be essential to monitor new data streams and continue to monitor old ones outside of epidemic periods to establish a baseline that can be used to calibrate signals of new outbreaks and estimate the exceedance caused by the ongoing transmission of novel pathogens (100).

**Conclusion**
Data and modeling needs change over the course of a pandemic and vary by the jurisdictional dimensions, requiring anticipatory, rapid, dynamic, and locally adapted and scaled activities to optimize pandemic management and population health. Here, we have sought to describe concepts, tools, and strategies to address those needs, building on those enacted during the COVID-19 pandemic and those that could have facilitated this work. While not a comprehensive list, we hope that the ideas we propose and envision serve as a useful resource and guide in efforts to manage ongoing infectious diseases challenges and preparedness for the inevitable next pandemic.

**E. References**

**BOX: Modeling and analytic support for STLT jurisdictions: the role of academic groups**

COVID-19 stimulated numerous collaborations between STLT health authorities and university and other research institutions to support decision making with modeling, analytics, and forecasts. These took multiple forms, ranging from the establishment of advisory councils to mayors and governors, to bilateral collaborations (101, 102)and formal consortia (https://modelingconsortium.ucsf.edu/). However, there are stark differences among jurisdictions in the number of such academic groups within the jurisdiction and/or with existing or prior links with the jurisdiction's health department, creating inequities in access to this kind of advice. The benefits of working with academic partners can include local knowledge and the capacity to surge efforts in an emergency. Potential barriers to such collaborations that should be addressed up front where possible include academics' need and incentives to publish, which may compete for time with their role in decision support, as well as the demands of academic schedules, whereby, for example, a key analyst on a project may have to devote effort to exams at times when they are needed for decision support. Administrative preparedness in the form of preexisting data use agreements can vastly accelerate these efforts.

Establishment of trust is essential to the success of academic – STLT collaborations. Elected and health officials at the symposium noted the repeated challenges of figuring out which models and modelers to trust, both locally and nationally. Participants observed that academic collaborations were most effective when there was a pre-existing relationship between the groups and the jurisdictions, and noted the benefits to both parties of cultivating these relationships in "peacetime" through collaboration on non-pandemic activities. Academic groups' ability to speak freely can lend credibility and objectivity to their analyses; however, trust can be undermined if academic groups with access to limited, publicly available data release analyses in publications or preprints that may be inconsistent with more complete data that are available to health departments but not publicly available. Frequent contact to share tentative conclusions and compare them against the evolving understanding of health officials can enhance the quality of analyses by incorporating more complete data, if these can be shared, and can enhance the trust between the parties, improving future interactions. When such interactions work well, they do not stifle the conclusions of academic groups but rather ensure that these conclusions are based on the best current understanding and to ensure that health officials are aware of what is being published about data from their jurisdictions. Academic incentives and structures are particularly not suited for routinely repeated analyses, such as reproductive number estimation, nowcasting, and forecasting, although academic centers have played key roles in these areas for over two years during COVID-19. Automation, as in the California consortium's dashboard, is one solution. CDC's Center for Forecasting and Outbreak Analytics is beginning to take on some of these tasks and will increasingly serve as a focal point for such repeated, real-time analyses.

**BOX: National insight from local evidence**

Implementation of public health policies is a state/territorial/tribal and local responsibility in the U.S., as we noted above. Infectious disease surveillance is also decentralized, often with two levels of reporting (local/county and state/territorial) below the national level. From the perspective of national decision makers seeking a clear picture of an unfolding pandemic, decentralized surveillance has obvious limitations, particularly in a setting where data systems and data use agreements vary across jurisdictions. Efforts are underway, and should be expanded, to improve the speed, completeness, and accuracy of data flowing from states, localities, and health care systems to the CDC and other federal actors. Such efforts are essential for timely situational awareness and for calibrating the outputs of scenarios and forecasts to granular (state or county-level) data to form a national picture.

While incidence, prevalence, and health care burden are intrinsically local quantities that need to be estimated everywhere and over time, many aspects of surveillance and associated epidemiology are generalizable, such that findings in one local jurisdiction can inform control measures everywhere. These include characteristics of the pathogen, such as severity and natural history; and characteristics of countermeasures, such as test sensitivity and the effectiveness of drugs and vaccines. For these purposes, local conditions can facilitate detailed characterization that may not be possible, but also may not be necessary, on a larger geographic scale.

Some of the earliest evidence of low severity for the 2009 H1N1 influenza pandemic came from a study at the University of Delaware, where a comparatively self-contained population could be studied in detail (103). We noted in a postmortem of that pandemic that the findings from that study were not widely known until months later because of limited dissemination (3), arguably prolonging the state of alarm unnecessarily during that pandemic. In COVID-19, early findings of asymptomatic/presymptomatic infection and likely transmission from studies in a nursing home and a cruise ship respectively (16, 104) were documented very early and widely disseminated, but still did not fully inform control measures.

In many other cases, detailed surveillance and epidemiology in local jurisdictions or health systems provided evidence of national and international importance. A few examples included:

- Evidence from the Yukon-Kuskokwim (Alaska) Health Corporation about the persistence of antigen test positivity 5 or more days after initial positive test or symptom onset during the early Omicron era (88)
- Evidence from the Kaiser Permanente Southern California health system about the relative clinical severity of Omicron BA.1 variant compared to Delta before it and BA.2 after (66)
- others

Each of these provided evidence that could be generalized beyond the location where it was generated, because it concerned generalizable features of the infection or countermeasures based on its biology. The degree to which these investigations informed policy and guidelines varied, indicating a need for a systematic approach to disseminating findings of wide importance and updating guidance in a way that reflects the totality of data.

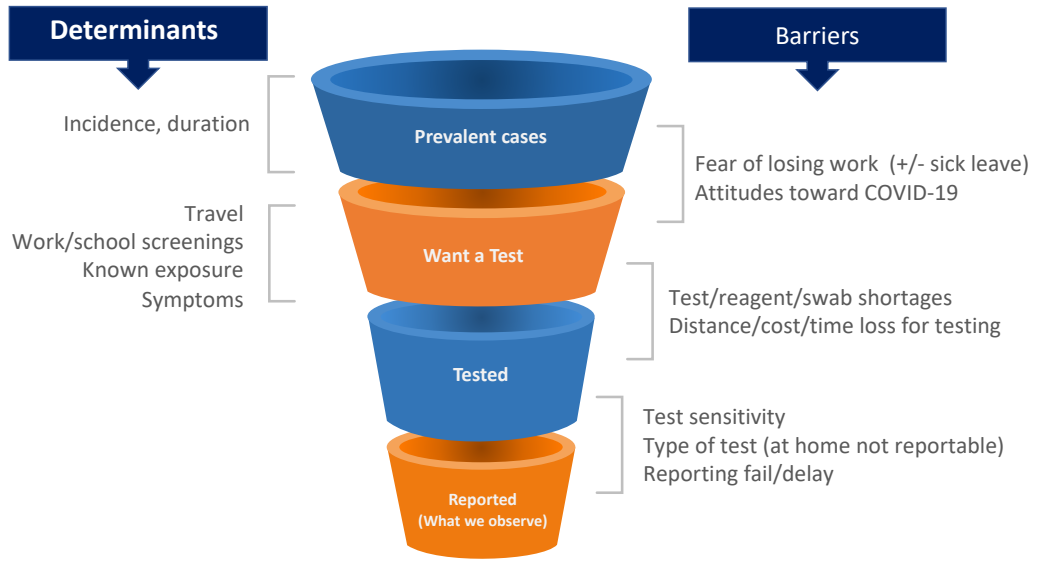

Fig. 1: Testing patterns that vary in space and time as a result of individual incentives (left) and barriers (right) determine a changing relationship between epidemiological quantities (top left) and reported case counts, making these counts an uncertain source of evidence for current case burden and for calibration of transmission models.

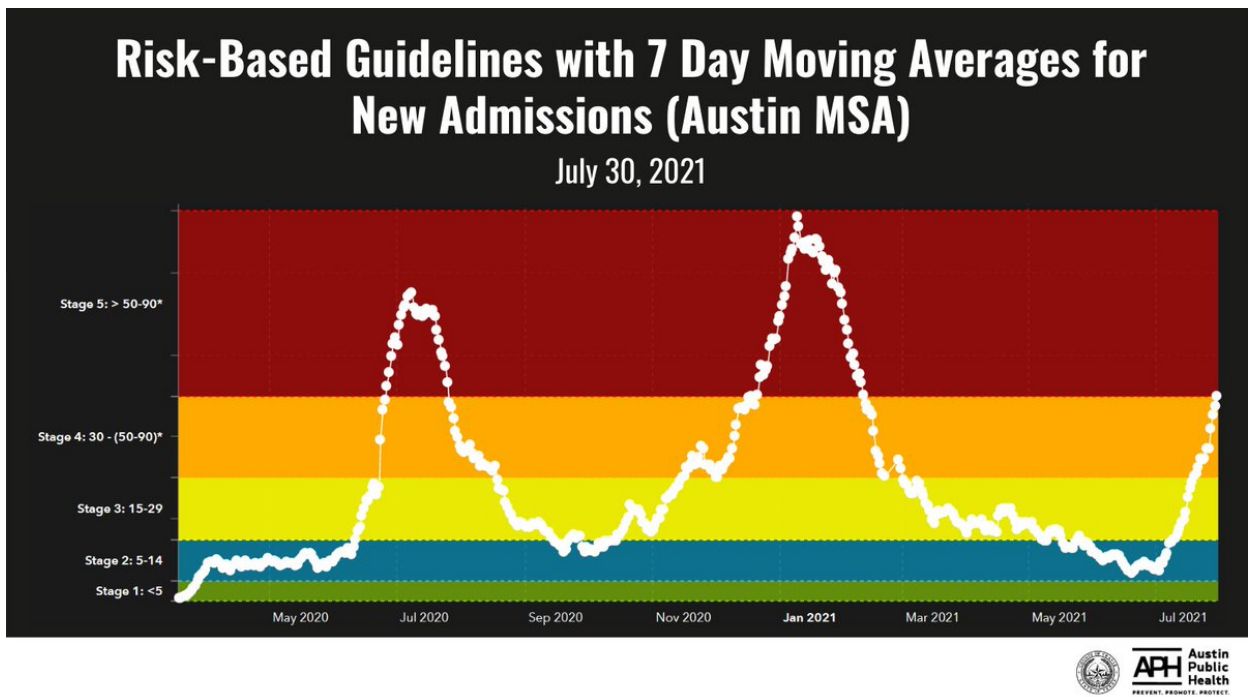

Figure 2: Evolution of COVID-19 risk levels through July, 2021 in Austin, Texas